\documentclass[]{emulateapj}
\usepackage{epsfig}

\def\CIV{\hbox{C~$\scriptstyle\rm IV\ $}}

\def\kms{\,{\rm km\,s^{-1}}}

\def\cmm{\,{\rm cm^{-2}}}

\def\msun{\,{\rm M_\odot}}

\def\Lya{Ly$\alpha\ $}
\def\etal{{et al.\ }}
\def\spose#1{\hbox to 0pt{#1\hss}}
\def\lta{\mathrel{\spose{\lower 3pt\hbox{$\mathchar"218$}}
     \raise 2.0pt\hbox{$\mathchar"13C$}}}
\def\gta{\mathrel{\spose{\lower 3pt\hbox{$\mathchar"218$}}
     \raise 2.0pt\hbox{$\mathchar"13E$}}}
\newcommand{\be}{\begin{equation}}
\newcommand{\ee}{\end{equation}}

\lefthead{Porciani \& Madau}
\righthead{Metals in the IGM}


\begin{document}

\title{The origin of intergalactic metals around Lyman-break galaxies}
\author{Cristiano Porciani\altaffilmark{1} and Piero Madau\altaffilmark{2}}
\altaffiltext{1}{Institute for Astronomy, ETH Z\"urich, 8093 Z\"urich, 
Switzerland; porciani@phys.ethz.ch.}
\altaffiltext{2}{Department of Astronomy \& Astrophysics, University of California, 
Santa Cruz, CA; pmadau@ucolick.org.}

\begin{abstract}
Theoretical and observational arguments suggest that 
the intergalactic medium (IGM) might have been polluted with metals produced
by early star formation.
In this scenario, Lyman-break galaxies (LBGs) at redshift $z=3$ 
are likely to be surrounded by old metal 
bubbles, relics of an era when the 
characteristic mass of galaxies was small and gas retainment more difficult.
We find that pregalactic enrichment of the IGM 
from $10^8-10^{10}\,\msun$ dwarf galaxies at $6<z<12$ 
can quantitatively explain the high cross-correlation 
between \CIV systems and LBGs observed at $z=3$.
The reason is twofold.
First, both LBGs and high-$z$ dwarfs are biased tracer of the mass
distribution and form from exceptionally high density fluctuations
which are strongly clustered.  
Second, the action of gravity tends to increase the spatial association
between metal bubbles and LBGs.
Our analysis shows that,
in order to match the abundance of \CIV systems observed at $z=3$, 
the metal bubbles generated by high-$z$ dwarfs
must have comoving sizes of  $\sim 100$ kpc.
We conclude that the observed galaxy--\CIV spatial 
association needs not to be generated by late ``superwinds'' from LBGs. 
\end{abstract}

\section{Introduction}

Quasar absorption lines constitute a fossil record that holds the clues to 
the exchange of mass, metals, and energy 
between active galactic nuclei, protogalaxies, and the intergalactic medium (IGM).
The distribution of metals in the IGM is
highly inhomogeneous, with a global cosmic abundance at $z=3$ of [C/H]$=-2.8\pm 
0.13$ for gas with overdensities $0.3<\delta<100$ (Schaye \etal 2003).
The epoch at which this cosmic enrichment occurred is uncertain, and so is the nature 
of the galaxies responsible for seeding the IGM with nuclear waste. The extent of 
metal pollution seem to require galaxy-scale outflows to overcome the 
gravitational potential of their host halos, and enriched material to be ejected 
far away from the density peaks where large galaxies form, gas cools, and 
star formation takes place.

Observationally, \CIV absorbers and Lyman-break galaxies (LBGs)
at $z\approx 3$ are found to be spatially correlated 
(Adelberger et al. 2003, hereafter ASSP).
The mean number of LBGs within a comoving distance $r$
from a detectable \CIV system ($N_{\rm CIV}\gta 10^{11.5}\,\cmm$)
equals one for $r=r_1=2.4\,h^{-1}$ Mpc.
This is much smaller than the value expected for a random distribution 
($r_1\simeq 3.9\,h^{-1}$ Mpc, ASSP).
Stronger \CIV systems seem to be even more closely associated with LBGs
($r_1\lta 0.5\,h^{-1}$ Mpc for $N_{\rm CIV}\gta 10^{13.5}\,\cmm$).
At the same time, the \Lya absorption produced by the IGM within $\sim 0.5\,h^{-1}$ Mpc
from an LBG appears to be systematically reduced with respect to random locations.
Moreover, high-resolution spectra reveal that LBGs have high star 
formation rates and strong winds with outflow velocities exceeding several hundred 
kilometers per second.
All these facts
led ASSP to argue that metal-rich ``superwinds'' from LBGs are responsible for 
distributing the product of stellar nucleosynthesis on Mpc scales.

In this {\it Letter} we argue that this interpretation may be an oversimplication of a far 
more complex physical picture. The starting point of our investigation is the 
fact that LBGs are highly biased tracers of the underlying mass-density field. 
The observed clustering amplitude
suggests that LBGs reside in dark-matter halos with mass 
$M>10^{11.5\pm 0.3}\,\msun$ at $z=2.9$ 
(Porciani \& Giavalisco 2002; Adelberger \etal 2005).
In hierarchical cosmogonies, 
these halos are expected to form within overdense regions (extending for a few Mpc) 
which are likely to be pre-enriched by subgalactic objects at much higher redshifts. 
The ejection of supernova debris from the shallow potential wells of collapsed, rare 
density fluctuations is, in fact, expected to lead to
an early era of metal pollution and preheating 
in biased regions of the IGM (e.g. Tegmark \etal 1993; Madau \etal 2001; Scannapieco 
\etal 2002; Thacker \etal 2002). 
We show that pregalactic enrichment from $10^8-10^{10}\,\msun$ dwarf galaxies at 
$6<z<12$ can quantitatively explain both the observed close spatial coincidence 
between \CIV systems and LBGs and the abundance of \CIV absorbers at $z=3$.
Throughout this paper we will adopt a flat $\Lambda$CDM background cosmology with 
parameters ($\Omega_\Lambda, \Omega_{\rm M}, \Omega_{\rm b}, n, \sigma_8, h)
=(0.3, 0.7, 0.05, 1, 0.87, 0.65)$.
 
\section{Basic theory}

The properties of the dark-matter halo population as a function of redshift 
can be approximately computed with the excursion-set formalism (Bond \etal 1991).
In this approach, the halo mass function 
$n(M,z)$ is determined by following the random walk of the linear density contrast, 
$\delta$, as a function of spatial resolution (parameterized by the mass variance, $S$)
and measuring the rate with which different trajectories first upcross a given 
threshold value, $t_c=\delta_{\rm c}(z)/D(z)$. Here $\delta_{\rm c}$ is the linear overdensity 
of a top-hat perturbation at collapse ($\delta_{\rm c}=1.686$ in a Einstein-de Sitter universe)
and $D(z)$ is the linear growth factor of density perturbations, 
normalized such that $D(z=0)=1$.

Porciani \etal (1998) have shown that the excursion-set formalism can be also used to 
derive the clustering properties of dark-matter halos
in the Lagrangian coordinate system (which gives the positions of proto-halo centers
in the linear density field, formally at $z\to \infty$).
This is obtained accounting for the spatial coherence of the linear density field
and simultaneously following a number of correlated random walks associated with different
spatial locations. 
More recently, Scannapieco \& Barkana (2002) generalized these results
to halos collapsing at different epochs. Here we use their analytical approximation 
to compute the Lagrangian correlation function
$\xi_{\rm L}(r_{\rm 
L}|M_1,z_1,M_2,z_2)$, between halos of mass $M_1$ forming at $z_1$ 
and halos of mass $M_2$ forming at $z_2>z_1$.
Hereafter, $M_1>10^{11.5} M_\odot$ will denote the mass of the host-halo of an LBG  
forming at $z_1=3$ while the free-parameters $M_2$ and $z_2$ 
will characterize subgalactic halos collapsing at earlier epochs.

In order to compare theory and observations, one needs to compute the halo clustering 
properties in the Eulerian coordinate system (in which galaxies are observed).
The mapping of halo positions from Lagrangian to Eulerian space 
depends on the gravity driven dynamics of the background mass density field
(Catelan et al. 1998).
Consider the evolution of a spherically-symmetric shell of Lagrangian radius $r_{\rm 
L}$ centered onto a halo of mass $M_1\gg M_2$. Its Eulerian position at redshift $z_1$
only depends on the mean overdensity $\Delta(z_1)$ within it: $r_{\rm E}=r_{\rm L}/[1+
\Delta(z_1)]^{1/3}$ (for the sheared case see Catelan et al. 1998). 
Using the spherical top-hat model it is possible to 
relate the non-linear density contrast $\Delta$ to its linearly extrapolated 
counterpart $\delta$, as $\delta\simeq \delta_c\,[1-(1+\Delta)^{-1/\delta_c}]/D(z)$
(Pavlidou \& Fields 2004). This implies
$r_{\rm E}=f(\delta,r_{\rm L})=r_{\rm L} [1-(D(z)\,\delta/\delta_c)]^{\delta_c/3}$.
%
%
What is the probability distribution of the linear overdensity within
a shell of given Lagrangian size centered onto an halo of mass $M_1$?
This can be easily computed using the excursion-set model. 
The conditional probability that a trajectory assumes the value $\delta$
when smoothed on the scale $r_{\rm L}$ (corresponding to the mass variance
$s$), given that it will first upcross the threshold $t_c(z_1)$ 
at the mass variance $S$ (corresponding to the mass $M_1$), is 
\be
{\cal P}(\delta,s|\delta_c,S)=\frac{{\cal W}(\delta,s)\,
{\cal P}_1(t_c-\delta,S-s)}
{{\cal P}_1(t_c,S)}, 
\label{cond}
\ee
where
\be
{\cal W}=
\frac{1}{\sqrt{2 \pi s}}\,\left[\exp\left(-\frac{\delta^2}{2s} 
\right)-\exp\left(-\frac{(2t_c-\delta)^2}{2s}\right)\right]
\ee
is the probability density of trajectories at $s$,
and
\be
{\cal P}_1(t_c,S)=\frac{t_c}{\sqrt{2 \pi S^3}}\,\exp
\left[\left(-\frac{t_c^2}{2 S} \right) \right]
\ee
denotes the probability distribution of first upcrossings.
The conditional probability distribution of $r_{\rm E}$ given a Lagrangian
separation can then be written as
\begin{eqnarray}
{\cal P}(r_{\rm E}|r_{\rm L})&=&
\int_{-\infty}^{t_c} {\cal P}(\delta|r_{\rm L}) 
\, \delta_D[r_{\rm E}-f(\delta,r_{\rm L})]\,d\delta \nonumber \\
&=&\frac{3\,{\cal P}(\bar{\delta}|r_{\rm L})}{r_{\rm L}\,D(z_1)\,|1-\bar{\delta}/t_c|^
{(\delta_c/3)-1}},
\label{pel}
\end{eqnarray}
where $\bar{\delta}=\delta_c\,[1-(r_{\rm E}/r_{\rm L})^{3/\delta_c}]/D(z_1)$,
${\cal P}(\delta|r_{\rm L})$ denotes the distribution given 
in equation (\ref{cond}), and $\delta_D(x)$ is the Dirac delta function.
%
%
\begin{figure}
\label{}
\plotone{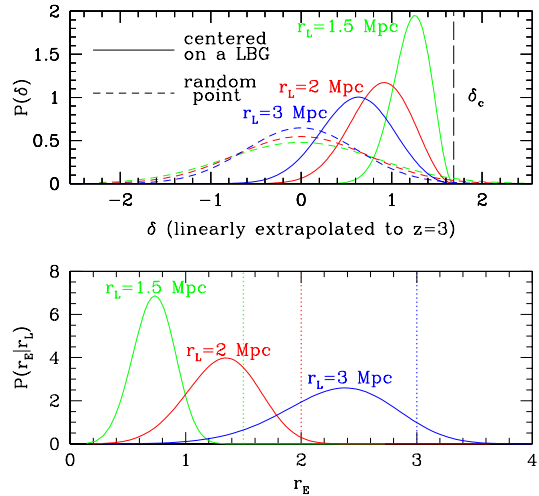}
\caption{
Top:
Probability distribution at $z=3$
of the linear overdensity in a sphere centered on
a dark matter halo with mass $M_1=10^{11.5} M_\odot$ ({\it solid}; see eq. 
\ref{cond}) and on a random point ({\it dashed}). The three sets of curves correspond
to spheres with Lagrangian radii of 1.5, 2 and 3 comoving Mpc. The Lagrangian radius
of the central halo is 1.3 Mpc.
Bottom: Probability distribution at $z=3$ of the Eulerian separation
between a dwarf galaxy (of any mass $M_2$ and collapse redshift $z_2$)
and an LBG (see eq. \ref{pel}) for a given Lagrangian 
separation $r_{\rm L}$ (indicated by labels and dashed lines).}
\end{figure}
The mean number of objects (per unit length) with mass $(M_2,M_2+dM_2)$ 
located at distance $r_{\rm E}$ from a halo of mass $M_1$ is then
\be
\frac{dN_2}{dr_{\rm E}}=4\pi\!\!\int_{R_1}^\infty \!\!\!\!\!
{\cal P}(r_{\rm E}|r_{\rm L})\,
n(M_2,z_2)\,[1+\xi_{\rm L}(r_{\rm L})]\,r_{\rm L}^2\,dr_{\rm L}\;,
\ee
with $R_1=(3 M_1/4 \pi \bar{\rho})^{1/3}$ the Lagrangian radius 
of the halo of mass $M_1$ and $\bar{\rho}$ the mean comoving density of the
universe.
For a population of halos with an extended range of masses at $z_1$, 
this becomes
\be
\left\langle \frac{dN_2}{dr_{\rm E}}\right\rangle_{M_1}= 
\frac{\int n(M_1,z_1)\,\displaystyle{\frac{dN_2}{dr_{\rm E}}}\,dM_1}{\int n(M_1,z_1) \,dM_1}\;.
\ee
Note that, by redshift $z_1$,
halos with mass $M_2$ and $r_{\rm L}<R_1$ have merged with the ``central'' 
object of mass $M_1$. 
The mass function of the remaining halos
is ${\cal N}(M_2,z_1)=
n(M_2,z_2)-\int n(M_1,z_1) N_{12}\,dM_1$, 
where $N_{12}=4\pi\,n(M_2,z_2)\, 
\int_{0}^{R_1} [1+\xi_{\rm L}(r_{\rm L})]\,r_{\rm L}^2\,dr_{\rm L}$ 
indicates the mean number of $M_2$ halos which ended up within
a single $M_1$ halo (this is in good agreement with Lacey \& Cole 1993).
From the definition of the Eulerian cross-correlation, 
we then obtain
\be
1+\xi(r_{\rm E})=\frac{1}{4\pi\,r_{\rm E}^2\,{\cal N}(M_2,z_1)}\left\langle \frac{dN_2}{dr_{\rm E}}\right\rangle_{M_1}\;.
\label{xieul}
\ee
%
%
\begin{figure*}
\label{vol_evo}
\plottwo{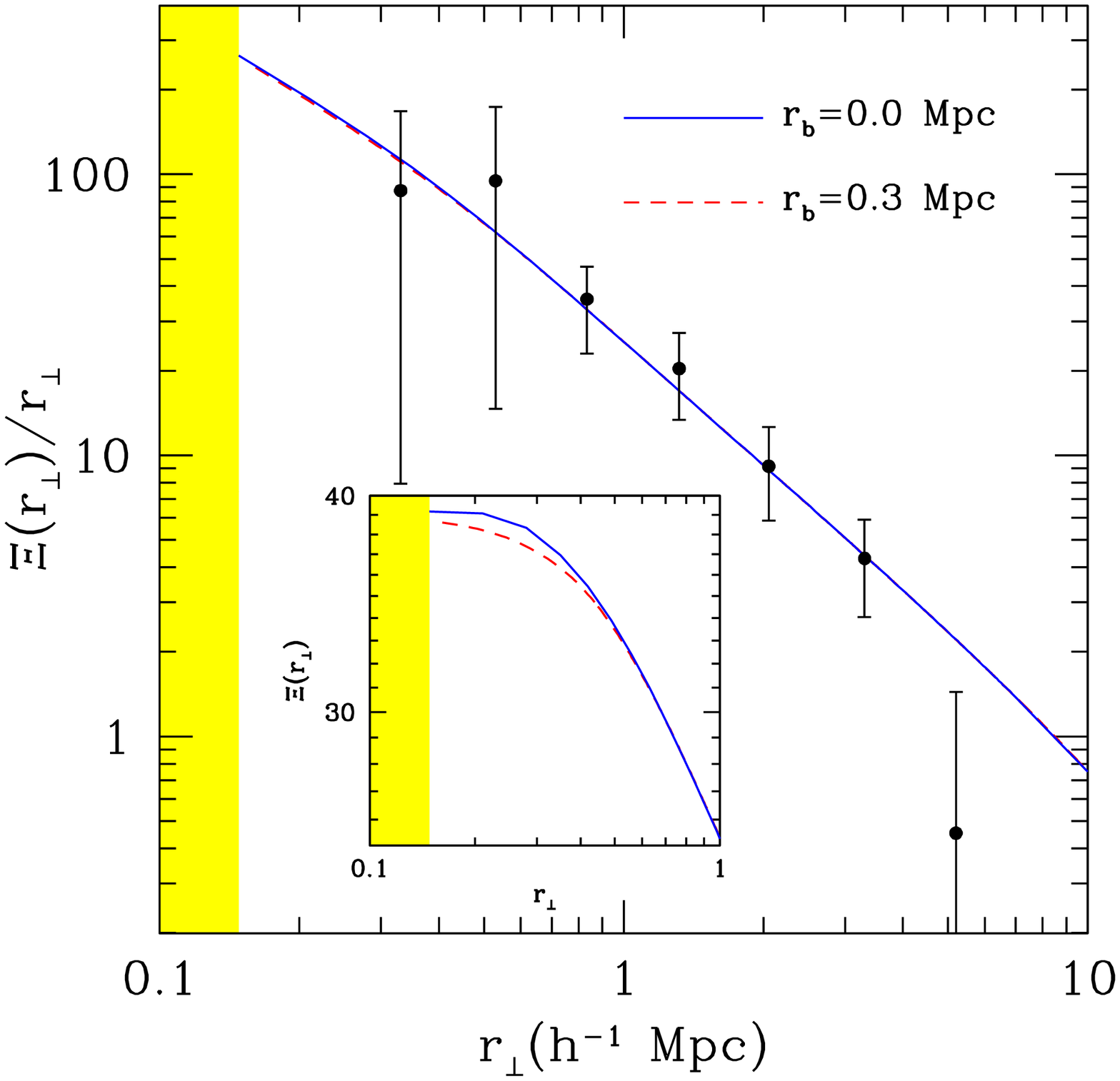}{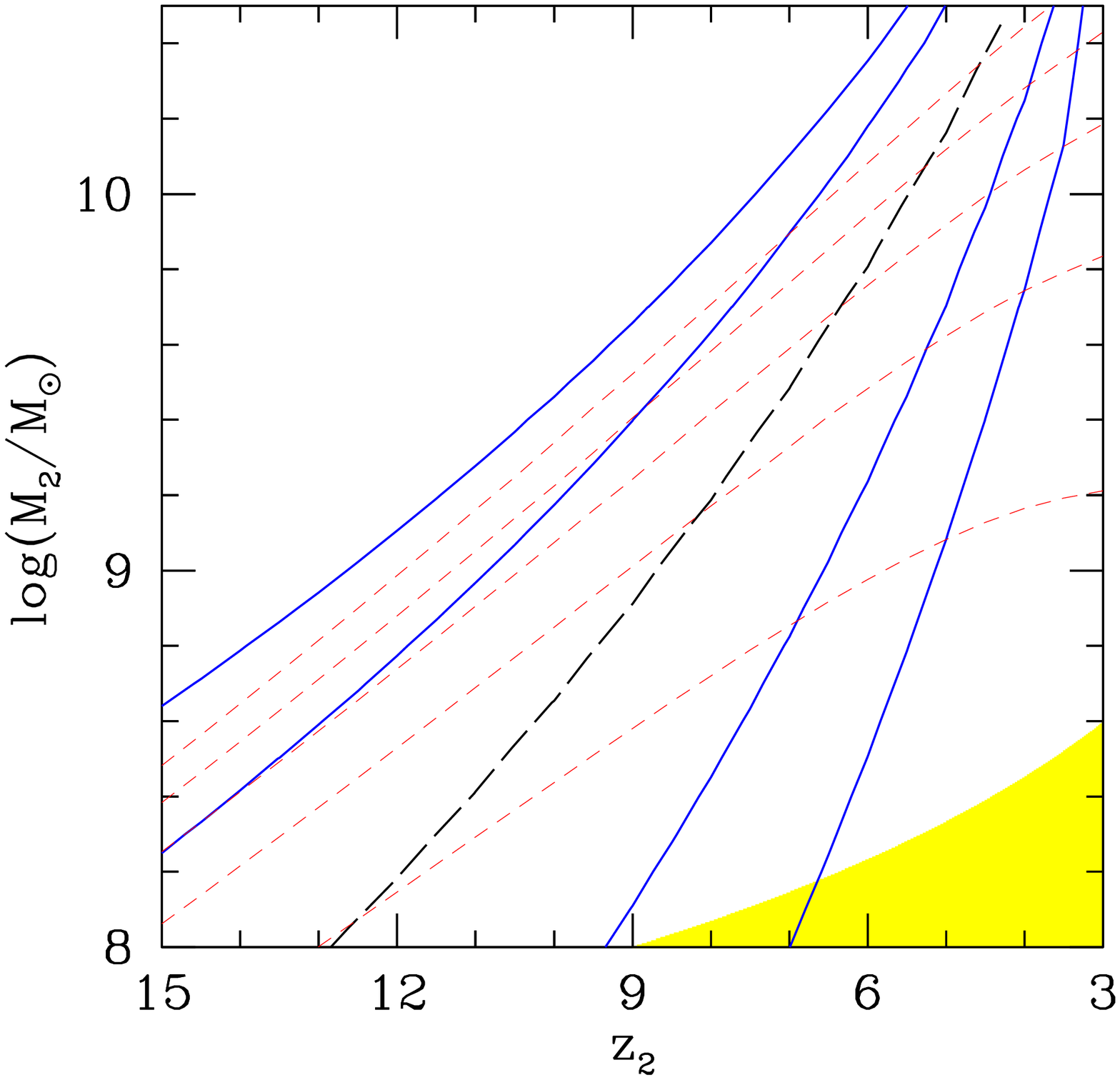}
\caption{
Left: Projected cross-correlation function between LBGs and \CIV systems.
Points with errorbars denote the observational data by ASSP,
while the solid line shows our best-fitting
model (which corresponds to $\chi^2_{\rm min}=3.6$). 
The shaded area indicates separations smaller than the Eulerian radius of a 
$10^{11.5} M_\odot$ halo,
where our model does not hold.
The inset shows the small-scale modifications due to a finite bubble size.
Right: Contour levels of the $\chi^2$ function 
obtained by fitting the data by ASSP
with the model described in the text.
{\it Solid contour}: $\Delta\chi^2=\chi^2-\chi^2_{\rm min}=2.3$ and 6.17
(respectively marking the 68.3 and 95.4 per cent confidence levels 
for two Gaussian variables). 
{\it Long-dashed curve:} set of points where 
$D(z_2)^{1.15}\,S(M_2)=2.9$,  
which describes the location of the degenerate minimum.
{\it Light, short-dashed lines:} set of points where the number density
of C IV systems matches the observed value for different values of $r_{\rm b}$
[ranging from 50 to 250 comoving kpc (from bottom to top) 
and spaced by 50 kpc]. 
The shaded area marks the region where virialized gas cannot cool via 
excitation of hydrogen Ly$\alpha$.
}
\end{figure*}

\section{Results}

We can now turn to the main point of this {\it Letter}. {\it Can the observed 
correlation between \CIV absorbers and LBGs at $z\approx 3$ be explained by 
SN-driven outflows from low mass systems at much higher redshifts}?  
Figure 1 shows that LBGs form within overdense regions extending for a few
comoving Mpc. Assume that these biased regions have been already polluted by 
protogalaxies, i.e. that \CIV systems are associated 
with metal ``bubbles'' that have been expelled from halos of mass $M_2$ at some epoch 
between $z_2$ and $z_1$.\footnote{Note that, at redshift $z_1$, most of the halos of 
mass $M_2$ have lost
their identity by merging into larger halos (of mass $<M_1$). We do not follow
the details of this process and assume that all the \CIV bubbles preserve
their identity. Since the bubble size is small with respect to the spatial 
separations achievable by observations, this assumption does not alter our 
conclusions.}
Most of the bubbles lying in the vicinity of the formation site of an LBG 
will be driven  by gravity towards the forming galaxy thus producing 
the observed spatial association (see bottom panel in Fig. 1).

In this scenario, the projected cross-correlation function between LBGs and 
the bubble centers is:
\be
\frac{\Xi(r_\perp)}{r_\perp}=\frac{2}{r_\perp}\,
\int_{r_\perp}^{\infty} \frac{r_{\rm E}\,\xi(r_{\rm E})}{(r_{\rm E}^2-r_\perp^2)^{1/2}}\, dr_{\rm E}\;,
\label{proj}
\ee
where $r_\perp$ denotes the component of the Eulerian separation $r_{\rm E}$ in
the plane of the sky.
For simplicity, we assume that each halo of mass $M_2$ 
gives origin to a spherical \CIV bubble of comoving radius $r_{\rm b}$. 
The cross-correlation function between LBGs and \CIV systems is then obtained
by convolving the correlation function in equation (\ref{proj}) with a 
2-dimensional top-hat window function
%
(see the Appendix in Porciani \& Giavalisco 2002).
In Figure 2, we show that the free parameters of our model can be tuned
to reproduce the correlation function measured by ASSP. 
Here we consider a unit logarithmic interval in mass for $M_1>10^{11.5} M_\odot$.
The best-fitting solution is not unique:
models with values of ($M_2$, $z_2$) linked
by the equation $D(z_2)^{1.15} S(M_2)=2.9$ reproduce the data 
with equal accuracy (right panel in Fig. 2). 
%
%
This corresponds to perturbations with $\nu\equiv t_c/\sqrt{S}\simeq 
D(z_2)^{-0.425}$. In the redshift interval $6<z_2<12$, $\nu$ ranges from 2 to 2.7,
i.e. with increasing $z_2$ rarer and rarer density fluctuations are required to pollute 
the surroundings of LBGs in order to match the observed $\Xi$. 
Note that 
massive LBGs at $z=3$ corresponds to $\nu=2$ peaks, i.e. {\it the dwarf galaxies that 
are the source of \CIV absorption around LBGs must have been equally or more biased 
tracers of the mass distribution than the LBGs themselves.}
These high-$\nu$ halos form first and, most likely, inhibit star formation
(and thus the production of new metal bubbles) in lower density peaks which
collapse at a more recent epoch (Thacker et al. 2002).

Figure 2 also shows that the clustering data by ASSP cannot constrain $r_{\rm b}$.
Winds from star-forming low-mass systems at $z\sim 10$ are expected to sweep up 
regions of the IGM of comoving size $r_{\rm b}<250\,$kpc (Mori \etal 2002; Furlanetto 
\& Loeb 2003) while $\Xi$ has been measured for separations $r>0.3\,h^{-1}$ Mpc
where the bubble size leaves no imprint.
We thus constrain $r_{\rm b}$ by comparing the observed mean number of \CIV 
absorbers per unit redshift with the expectations from our model.
From Songaila (2001), we infer that $dN/dz\simeq 8$ for 
\CIV systems with $N_{\rm CIV}>10^{13}\,\cmm$ at $z\sim 3.2$. 
This corresponds to a bubble size of the order of 100 comoving kpc and 
implies that $r_{\rm b}$ has to increase when one moves towards lower redshifts 
along the $(M_2,z_2)$ degeneracy line (see Fig. 2).
Thus enriched outflows of size 
$r_{\rm b}=100\,$kpc around $10^{9}\,\msun$ halos at $z=9$ 
(or of size $r_{\rm b}=200\,$kpc around $10^{10}\,\msun$ halos at $z=6$)
might explain 
both the mean absorption-line density 
and the LBG--\CIV cross-correlation observed at redshift 3.

\section{Discussion}

In hierarchical cosmologies, LBGs represent only the tail end of a 
larger population of pregalactic starbursts that formed at much higher redshifts.
The metal pollution of the IGM may then have started at early times, perhaps by the same
sources responsible for hydrogen reionization. 
%
In ``pregalactic enrichment'' scenarios, LBGs are expected to form
within biased regions of the IGM (as indicated by their strong clustering), 
which are the sites of previous star formation and enrichment. 
In this {\it Letter} we showed 
that the LBG--\CIV cross-correlation function
measured by ASSP can be explained by outflows from dwarf galaxies 
($10^8\lta M\lta 10^{10}\,\msun$) collapsing at $6<z<12$.
These originate from density peaks with $2<\nu<2.7$ (equally or more biased than the 
LBGs themselves) and are gravitationally pulled towards the LBG formation sites.
Note that, assuming that LBGs are hosted by halos with $M>10^{11.5}\,\msun$,
our model automatically reproduces also the observed correlation
function of LBGs (Porciani \& Giavalisco 2002; Adelberger \etal 2005). 
Contrary to previous beliefs,
our results then indicate that the observed coincidence between LBGs 
and \CIV systems does not necessarily imply that the metals observed in the IGM at 
$z\sim 3$ were driven out of massive LBGs by a superwind. 

Typical LBGs seem to generate gas outflows 
with velocities in the range $0<v<600\,\kms$ (ASSP). 
Material moving at a constant speed of 
$600\,\kms$ for 300 Myr can only cross 0.7 comoving Mpc at $z=3$. 
Realistic SN-driven winds will slow down as they physically displace baryons in 
the IGM, and can propagate out to $r\lta 1\,$Mpc only if they are able to tap 100\% 
of the available SN energy (e.g. Croft \etal 2002). 
The association between LBGs and the weakest 
metal line absorbers cannot then be readily explained by superwinds. Moreover, while
the distribution of heavy elements in the \Lya forest shows a positive gradient with 
overdensity,  there is little evolution of the total metal content of the IGM over the 
redshift range $z=$1.5--5 (Schaye \etal 2003; Pettini \etal 2003; Songaila 2001), as 
expected if most of the intergalactic metals were already in place at the highest 
currently observable redshifts.
All this provides supporting evidence for our proposed scenario where
SN-driven winds from a biased population of subgalactic halos at very high redshift 
pollute the regions surrounding LBGs with metals.
This does not require the winds to propagate to large distances.
Comoving metal bubbles of $\sim 100$ kpc expelled by halos of $M_2=10^{9}\,M_\odot$
at $z=9$ are sufficient to explain the observed abundance of \CIV systems at $z\sim 3$.

It is fair to point out at this stage that the initial epochs of the galaxy formation 
process are currently only poorly understood. N-body$+$hydrodynamical simulations 
of CDM cosmogonies have convincingly 
shown that the IGM is expected to fragment into bound structures at early times, 
but are much less able to predict the efficiency with which 
gravitationally collapsed objects lit up, reionized, and enriched the universe at 
the end of the ``dark ages''. 
Still, from the arguments presented above, it would 
seem unwise to use the spatial association between \CIV absorbers and LBGs 
and draw far-reaching conclusions on the extent of galactic superwinds and the direct 
impact of LBGs on the \Lya forest. Uncovering the relative contributions of LBGs and 
young dwarfs to the pollution of the IGM may prove difficult. Metals ejected during
an early episode of star formation may be seen in absorption against bright background
sources such as rare quasars or gamma-ray bursts at $z\sim 10$
(Furlanetto \& Loeb 2003). 
This, however, requires high-resolution spectroscopy of faint sources in the near-IR
and may have to wait for the {\it James Webb Space Telescope}. 
On the other hand, old wind bubbles
may have different abundance patterns than young winds from LBGs.
As recently pointed out by Cen \etal (2004), the gaseous relics
of an era when the characteristic mass of galaxies was small, gas retainment more 
difficult, and metal recycling more limited, are expected to contain  
lower ratios of secondary (e.g. Nitrogen) to primary 
(e.g. Carbon, Oxygen) metals.
For instance, theoretical yields from pair-instability SNe originating from massive
Population III stars are characterized by large Si/C ratios (Heger \& Woosley 2002).
Tests relying on measurements of relative metal abundances at high-$z$, however, are 
made impractical by the uncertainties in the ionization corrections and 
the lack of information regarding the stellar initial mass function.

\acknowledgments
We thank Evan Scannapieco for providing
an early set of Lagrangian correlations in tabulated form and 
an updated version of the Gemini code. 
Support for this work was provided by NASA grants NAG5-11513 and NNG04GK85G, and by 
NSF grant AST-0205738 and PHY99-07949 (PM).

{}

\end{document}